\documentclass[pra,twocolumn,floatfix,a4paper,superscriptaddress]{revtex4}

\usepackage{bm,color,amsmath,txfonts}
\usepackage{graphicx}
\usepackage{siunitx}
\usepackage{subfigure}
\usepackage{verbatim}
\usepackage{dcolumn}
\usepackage{bm}
\usepackage{epsf}
\usepackage{xcolor}
\usepackage{hyperref}
\usepackage{hhline}
\usepackage{braket}
\usepackage{float}
\usepackage{enumerate}
\usepackage{bbm}
\usepackage{lipsum}
\usepackage{appendix}

\begin{document}

	\title{Entangling ferrimagnetic magnons with an atomic ensemble via opto-magnomechanics}	
	
	\author{Zhi-Yuan Fan}
	\affiliation{Interdisciplinary Center of Quantum Information, State Key Laboratory of Modern Optical Instrumentation, and Zhejiang Province Key Laboratory of Quantum Technology and Device, School of Physics, Zhejiang University, Hangzhou 310027, China}
	\author{Hang Qian}
	\affiliation{Interdisciplinary Center of Quantum Information, State Key Laboratory of Modern Optical Instrumentation, and Zhejiang Province Key Laboratory of Quantum Technology and Device, School of Physics, Zhejiang University, Hangzhou 310027, China}
	\author{Xuan Zuo}
	\affiliation{Interdisciplinary Center of Quantum Information, State Key Laboratory of Modern Optical Instrumentation, and Zhejiang Province Key Laboratory of Quantum Technology and Device, School of Physics, Zhejiang University, Hangzhou 310027, China}
	\author{Jie Li}\thanks{jieli007@zju.edu.cn}
	\affiliation{Interdisciplinary Center of Quantum Information, State Key Laboratory of Modern Optical Instrumentation, and Zhejiang Province Key Laboratory of Quantum Technology and Device, School of Physics, Zhejiang University, Hangzhou 310027, China}

\begin{abstract}
We show how to prepare macroscopic entanglement between an atomic ensemble and a large number of magnons in a ferrimagnetic YIG crystal. Specifically, we adopt an opto-magnomechanical configuration where the magnetostriction-induced magnomechanical displacement couples to an optical cavity via radiation pressure, and the latter further couples to an ensemble of two-level atoms that are placed inside the cavity. We show that by properly driving the cavity and magnon modes, optomechanical entanglement is created which is further distributed to the atomic and magnonic systems, yielding stationary entanglement between atoms and magnons. The atom-magnon entanglement is a result of the combined effect of opto- and magnomechanical cooling and optomechanical parametric down-conversion interactions.  A competition mechanism between two mechanical cooling channels is revealed. We further show that genuine tripartite entanglement of three massive subsystems, i.e., atoms, magnons and phonons, can also be achieved in the same system. Our results indicate that the hybrid opto-magnomechanical system may become a promising system for preparing macroscopic quantum states involving magnons, photons, phonons and atoms.
\end{abstract}
	\date{\today}
	\maketitle

Cavity optomechanics (COM) explores the interaction between the electromagnetic field and mechanical motion via radiation pressure~\cite{RMP2014}. The past decade has witnessed significant progress in the field of COM in experimentally preparing macroscopic quantum states of massive mechanical oscillators, including the realization of entanglement between a mechanical oscillator and an electromagnetic field~\cite{Palomaki}, entanglement between two mechanical oscillators~\cite{mechanical1,mechanical2,mechanical3}, and quantum squeezing of mechanical motion~\cite{qusqz}, etc. 
	
In analogy to cavity optomechanics, cavity magnomechanics (CMM)~\cite{16SciAdv,JieLi18,PRX21,RCShen2022} has recently received increasing attention because of its potential for preparing quantum states at larger scales~\cite{JieLi18,JL19PRA,JL19NJP,Tan19PRre,Ding,JLQST21,OE21,ChenPRA21,JLNSR22,HussainPRA22,QiuPRA22,HQianQST23,Asjad23}, as well as its various promising applications in quantum information science and quantum technologies~\cite{SQapp1,SQapp2,SQapp3,Ding19,JieLi20,SQapp5,SQapp6,SQapp7,SQapp8,FanPRA21,SQapp10,SQapp11}. It studies interactions between microwave cavity photons, magnons (quanta of the spin wave), and magnetostriction-induced vibration phonons in magnetically ordered materials, such as yttrium-iron-garnet (YIG)~\cite{16SciAdv,JieLi18,PRX21,RCShen2022}. The combination of COM and CMM, realized by coupling the magnomechanical displacement to an optical cavity via radiation pressure, forms the new system of opto-magnomechanics (OMM)~\cite{FanPRA21,QST23,Fanarxiv}. Such a hybrid system enables us to optically read out magnon population~\cite{FanPRA21} and arbitrary magnonic quantum states in solids~\cite{QST23}, and prepare optomagnonic~\cite{QST23} and microwave-optics~\cite{Fanarxiv} entanglement. Thus, the system would find promising applications in quantum information processing and quantum networks.

\begin{figure}[b]
		\includegraphics[width=1\linewidth]{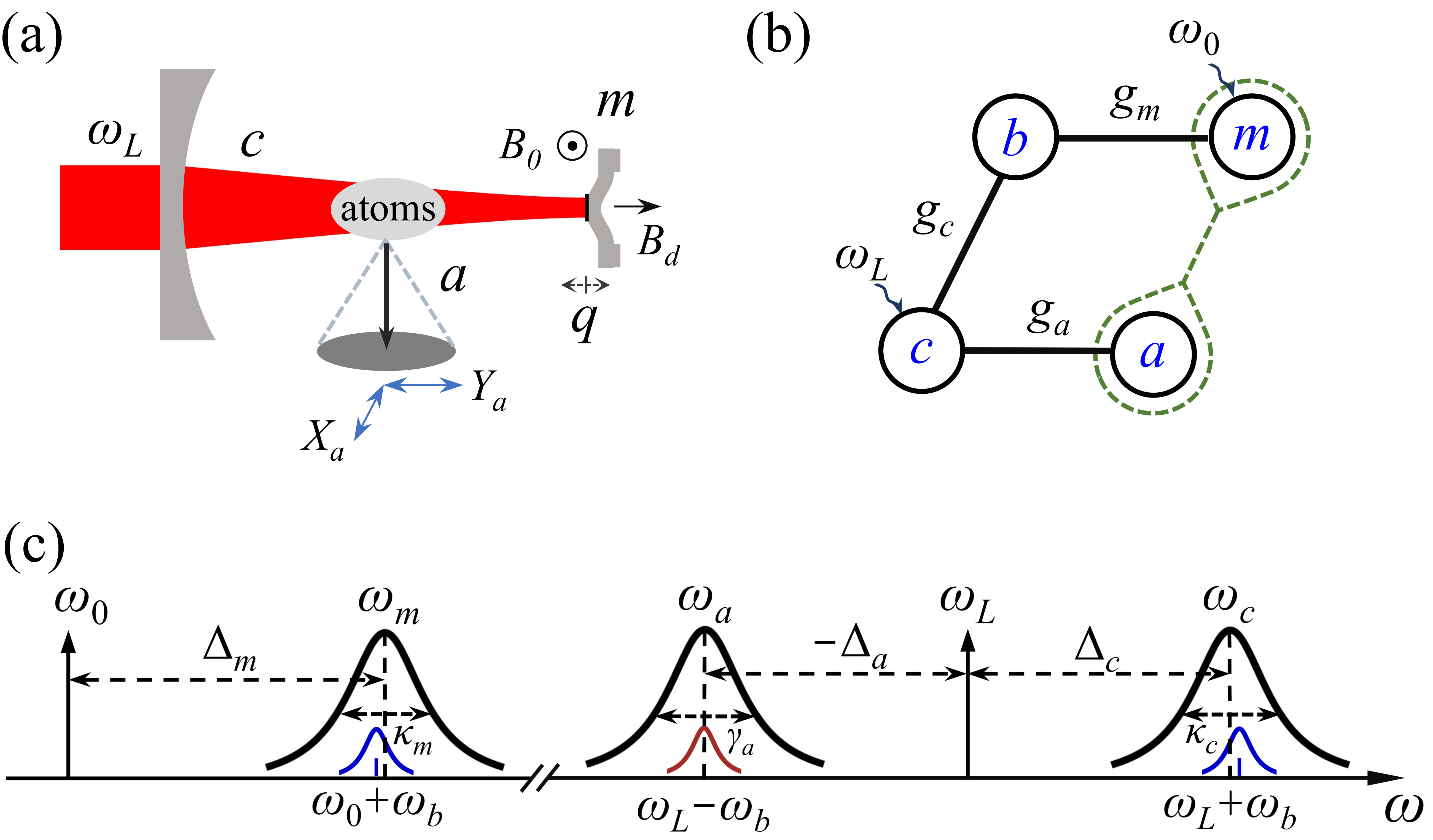}
		\caption{(a)-(b) Sketch of the atom-OMM system. An optical cavity mode ($c$) driven by a laser at frequency $\omega_L$ couples to an ensemble of two-level atoms ($a$) and a magnon mode ($m$) in a YIG crystal by the mediation of a mechanical vibration mode ($b$) induced by magnetostriction. (c) Mode frequenies and linewidths adopted in the protocol. In the resolved sideband limit ($\omega_b\gg \kappa_{c,m}$), when the cavity and atomic frequencies match the anti-Stokes sideband at $\omega_L+\omega_b$ and the Stokes sideband at $\omega_L-\omega_b$ of the laser drive field, respectively, and when further the magnon mode is resonant with the anti-Stokes sideband at $\omega_0+\omega_b$ of the microwave drive field, stationary entanglement between atoms and magnons is established. }
		\label{fig1}
	\end{figure}

In this article, we show how to prepare macroscopic entanglement between ferrimagnetic magnons and an atomic ensemble by using such a novel OMM system. Specifically, we further couple the optical cavity of the OMM to an ensemble of two-level atoms that are initially prepared in their ground state.   Keeping the atoms and magnons in the low-excitation limit, which allows us to bosonize the atomic polarization and the collective spins, the system then becomes a four-mode bosonic system. By strongly driving the cavity with a red-detuned laser, the mechanical motion scatters the driving photons onto two sidebands. When the cavity resonance and atomic frequencies match the anti-Stokes and Stokes sidebands, respectively, an entangled state between atoms and vibration phonons can be created. By further driving the magnons with a relatively weak red-detuned microwave field, which activates the magnon-phonon state-swap interaction, the atom-phonon entanglement is then partially distributed to magnons, yielding a macroscopic entangled state of magnons and atoms. The entangled state is stationary and robust against bath temperature. A strategy is provided to detect the entanglement.

The system we consider is depicted in Fig.~\ref{fig1}(a), which consists of optical cavity photons, magnons in a YIG crystal, magnetostriction-induced vibration phonons, and an ensemble of  two-level atoms.  The magnomechanical displacement couples to the optical cavity via radiation pressure (a dispersive interaction), e.g., by attaching a small highly reflective mirror pad onto the surface of a YIG micro bridge~\cite{QST23,bridge,Fanarxiv}. The YIG bridge is micron-sized and can support long-lived spin-wave excitations with the frequency in gigahertz and mechanical vibration phonons with the frequency ranging from tens to hundreds of megahertz~\cite{bridge}. The large frequency mismatch of the magnon and mechanical modes permits a dominant magnon-phonon dispersive coupling~\cite{QST23}. Note that the attached mirror pad should be fabricated sufficiently small compared to the YIG bridge, such that there is negligible bending displacement (relative motion), and the YIG bridge and the attached mirror can stick together tightly, which oscillate approximately with the same frequency.   Alternatively, one may adopt the `membrane-in-the-middle' configuration~\cite{Jack} by placing the YIG bridge in the middle of the optical cavity, which can also realize the dispersive coupling between the magnomechanical displacement and the optical cavity.

The Hamiltonian of the system is given by
	\begin{equation} \label{hamiltonian}
		\begin{split}
			H/\hbar=&\ \omega_c c^\dagger c+\omega_m m^\dagger m+\frac{\omega_a}{2}S_z+\frac{\omega_b}{2}\left(q^2+p^2\right)\\  
			&+g_a \left(S_+c +S_- c^\dagger \right)- g_c c^\dagger cq+ g_m m^\dagger mq+H_\mathrm{dri}/\hbar,
		\end{split}
	\end{equation}
where $c$ ($c^\dagger$) and $m$ ($m^\dagger$) are the annihilation (creation) operators of the cavity and magnon modes, respectively, satisfying $[k,k^{\dag}]=1$ ($k=c,m$). $q$ and $p$ ($[q,p]=i$) are the dimensionless position and momentum of the mechanical mode.  The collective spin operators of an ensemble of $N_a$ two-level atoms with natural frequency $\omega_a$, $S_{\pm,z}=\Sigma_{i=1}^{N_a}\sigma^{(i)}_{\pm,z}$, with $\sigma_{\pm,z}$ being the Pauli matrices, which satisfy the commutation relations $[S_+,S_-]=S_z$ and $[S_z,S_\pm]=\pm 2S_\pm$.  $\omega_c$, $\omega_m$ and $\omega_b$ are the resonance frequencies of the cavity, magnon and mechanical modes, respectively, and the magnon frequency can be adjusted in a large range by varying the strength of the bias magnetic field $B_0$. The atom-cavity coupling strength $g_a=\mu \sqrt{\omega_c/2\hbar\epsilon_0 V_c}$, with $\mu$ being the atomic dipole moment, $V_c$ the cavity mode volume, and $\epsilon_0$ the vacuum permittivity. $g_c$ ($g_m$) denotes the bare optomechanical (magnomechanical) coupling strength, which can be greatly enhanced by strongly driving the cavity (magnon) mode. The last term is the driving Hamiltonian, $H_{\mathrm{dri}}=i\hbar E(c^\dagger e^{-i\omega_L t}-\mathrm{H.c.})+i\hbar \Omega_d(m^\dagger e^{-i\omega_0 t}-\mathrm{H.c.})$, where $E=\sqrt{2\kappa_c P_L/\hbar\omega_L}$ represents the coupling strength between the cavity and the laser drive field, with $P_L$ ($\omega_L$) being the power (frequency) of the laser, and $\kappa_c$ the cavity decay rate. The Rabi frequency $\Omega_d=\frac{\sqrt{5}}{4}\gamma\sqrt{N}B_d$~\cite{JieLi18} denotes the coupling between the magnon mode and the drive magnetic field with amplitude $B_d$ and frequency $\omega_0$, and $\gamma$ is the gyromagnetic ratio and $N$ is the number of spins in the YIG crystal.  Note that the magnons are assumed in the low-excitation limit, $\langle m^\dagger m \rangle \ll 2Ns$, where $s=\frac{5}{2}$ is the spin number of the ground state ${\rm Fe}^{\rm 3+}$ ion in YIG~\cite{JieLi18}. This validates the bosonic description of the spins, where the system can be well described by a harmonic oscillator~\cite{SQapp1}.

The dynamics of the system governed by the Hamiltonian \eqref{hamiltonian} is generally complicated. It, however, can be simplified by assuming the atoms in the low-excitation limit, where the excitation probability of a single atom is small~\cite{DV08}. We assume that the atoms are initially prepared in their ground state, so that $S_z \simeq \langle S_z \rangle \simeq -N_a$, and are {\it off-resonantly} coupled to the optical cavity. In this case, the dynamics of the atomic polarization can be described by bosonic operators. Specifically, the atomic annihilation operator can be defined as $a=S_-/\sqrt{|\langle S_z \rangle|}$, and it satisfies the bosonic commutation relation $[a,a^{\dag}]=1$.  Consequently, we obtain the fully bosonized Hamiltonian, given by
	\begin{equation} 
		\begin{split}
			H/\hbar=&\sum_{j=a,c,m} \omega_j j^\dagger j+\frac{\omega_b}{2}\left(q^2+p^2\right)+H_\mathrm{dri}/\hbar\\  
			&+g_N \left(a^\dagger c +a c^\dagger \right)- g_c c^\dagger cq+ g_m m^\dagger mq,
		\end{split}
	\end{equation}
where $g_N=g_a\sqrt{N_a}$ is the effective atom-cavity coupling strength. 

By including dissipation and input noise of each mode and working in the interaction picture with respect to $\hbar\omega_L(a^\dagger a+c^\dagger c)+\hbar\omega_0 m^\dagger m$, we obtain the following quantum Langevin equations (QLEs) of the system:
\begin{equation}  \label{QLEs1}
		\begin{split}
			\dot{a}=&-i\Delta_a a-\gamma_a a-ig_N c+\sqrt{2\gamma_a}a_{\rm in}, \\
			\dot{c}=&-i\Delta_c c-\kappa_c c+ig_c cq-ig_N a+E+\sqrt{2\kappa_c}c_{\rm in},  \\
			\dot{m}=&-i\Delta_m m-\kappa_m m-ig_m mq+\Omega_d+\sqrt{2\kappa_m}m_{\rm in}, \\
			\dot{q}=&\ \omega_b p,\,\,\,\,\, \dot{p}=-\omega_b q-\gamma_b p+g_c c^\dagger c-g_m m^\dagger m+\xi, 
		\end{split}
	\end{equation}
where $\Delta_{a(c)}=\omega_{a(c)}-\omega_L$ and $\Delta_m=\omega_m-\omega_0$. $\gamma_{a}$ is the decay rate of the atomic excited level and $\gamma_{b}$ ($\kappa_{m}$) is the dissipation rate of the mechanical (magnon) mode. $j_{\rm in}(t)$ ($j=a,c,m$) denote the zero-mean input noise operators, which obey the following correlation functions $\langle j_{\rm in}^\dagger(t) j_{\rm in}(t')\rangle=N_j(\omega_j)\delta(t-t')$ and $\langle j_{\rm in}(t) j^\dagger_{\rm in}(t')\rangle=[N_j(\omega_j)+1]\delta(t-t')$. $\xi(t)$ is the Hermitian Brownian noise operator acting on the mechanical oscillator, which is intrinsically non-Markovian, but a Markovian approximation can be taken for a large mechanical quality factor $Q_b=\omega_b/\gamma_b\gg 1$~\cite{DV01}. In this case, $\xi(t)$ takes a $\delta$-autocorrelation: $\langle \xi(t)\xi(t')+\xi(t')\xi(t) \rangle/2\simeq \gamma_b[2N_b(\omega_b)+1]\delta(t-t')$. Here, $N_k(\omega_k)=[\mathrm{exp}(\hbar\omega_k/k_B T)-1]^{-1}$ ($k=a,c,m,b$) are the mean thermal excitation number of each mode at bath temperature $T$, with $k_B$ as the Boltzmann constant.  

The creation of strong quantum correlations, like entanglement, in the system requires sufficiently strong opto- and magnomechanical interactions. To this end, we drive the cavity (magnon) mode with a strong laser (microwave) field, which leads to large steady-state amplitudes $|\langle c \rangle|$, $|\langle m \rangle|\gg 1$. This allows us to linearize the nonlinear opto- and magnomechanical dynamics around the steady state, which is implemented by writing each mode operator as the sum of its classical average and quantum fluctuation operator, i.e., $k=\langle k\rangle+\delta k$, and neglecting small second-order fluctuation terms. 
We aim to study quantum correlations among the macroscopic subsystems, and thus we focus on the dynamics of the quantum fluctuations (around the steady-state averages). The linearized QLEs describing the quantum fluctuations $(\delta x_a,\delta y_a,\delta x_c, \delta y_c,\delta q,\delta p,\delta x_m,\delta y_m)$, where $\delta x_j=(\delta j+ \delta j^\dagger)/\sqrt{2}$ and $\delta y_j=i(\delta j^\dagger-\delta j)/\sqrt{2}$ ($j=a,c,m$) denote the fluctuations of the amplitude and phase quadratures of the corresponding mode,  can be written in the matrix form of
\begin{equation} \label{Eqmatrix}
		\dot{u}(t)=A u(t)+n(t),
\end{equation}
where $u=\left(\delta x_a,\delta y_a,\delta x_c, \delta y_c,\delta q,\delta p,\delta x_m,\delta y_m \right)^{\rm T}$ and $n=\big(\sqrt{2\gamma_{a}}x_a^{\rm in},\sqrt{2\gamma_{a}}y_a^{\rm in},\sqrt{2\kappa_c}x_c^{\rm in},\sqrt{2\kappa_c}y_c^{\rm in},0, \xi,\sqrt{2\kappa_m}x_m^{\rm in},\sqrt{2\kappa_m}y_m^{\rm in}\big)^{\rm T}$, and the drift matrix $A$ is given by
\begin{equation} \label{drift}
		A=\begin{pmatrix}
			-\gamma_a & \Delta_a & 0 & g_N & 0 & 0 & 0 & 0\\
			-\Delta_a & -\gamma_a & -g_N & 0 & 0 & 0 & 0 & 0\\
			0 & g_N & -\kappa_c & \tilde{\Delta}_c & G_c & 0 & 0 & 0\\
			-g_N & 0 & -\tilde{\Delta}_c & -\kappa_c & 0 & 0 & 0 & 0\\
			0 & 0 & 0 & 0 & 0 & \omega_b & 0 & 0\\
			0 & 0 & 0 & -G_c & -\omega_b & -\gamma_b & 0 & G_m\\
			0 & 0 & 0 & 0 & -G_m & 0 & -\kappa_m & \tilde{\Delta}_m\\
			0 & 0 & 0 & 0 & 0 & 0 & -\tilde{\Delta}_m & -\kappa_m
		\end{pmatrix},
\end{equation}
where the effective detunings $\tilde{\Delta}_c=\Delta_c-g_c\langle q\rangle$ and $\tilde{\Delta}_m=\Delta_m+g_m\langle q\rangle$, which include the frequency shifts due to the mechanical displacement $\langle q\rangle=\big(g_c |\langle c \rangle|^2-g_m |\langle m \rangle|^2 \big)/\omega_b$, jointly caused by the opto- and magnomechanical interactions. The effective opto- and magnomechanical coupling strengths are $G_c=i\sqrt{2}g_c\langle c\rangle$ and $G_m=i\sqrt{2}g_m\langle m\rangle$, which are significantly enhanced by the large amplitudes of the cavity and magnon modes:
	\begin{equation} \label{mean}
		\begin{split}
			\langle m\rangle= \frac{\Omega_d}{\kappa_m+i\tilde{\Delta}_m},\,\,\,\,\,\,  \langle c\rangle = \frac{E(\gamma_a+i\Delta_a)}{g_N^2+(\gamma_a+i\Delta_a)(\kappa_c+i\tilde{\Delta}_c)}.
		\end{split}
	\end{equation}
The amplitude of the atomic mode can be obtained by $\langle a\rangle =-i g_N \langle c\rangle/(\gamma_a + i\Delta_a)$.   It should be noted that the drift matrix $A$ in Eq.\eqref{drift} is derived under the condition that $|\Delta_a|, |\tilde{\Delta}_c|, |\tilde{\Delta}_m| \gg \gamma_a, \kappa_c, \kappa_m$. This yields simpler approximate expressions $\langle m\rangle \simeq-i\Omega_d/\tilde{\Delta}_m$ and $\langle c\rangle \simeq iE \Delta_a/(g_N^2-\Delta_a\tilde{\Delta}_c)$, which are pure imaginary numbers and thus give rise to approximately real couplings $G_m$ and $G_c$. In fact, as will be shown later, the condition $|\Delta_a|, |\tilde{\Delta}_c|, |\tilde{\Delta}_m| \simeq \omega_b \gg \gamma_a, \kappa_c, \kappa_m$ (c.f. Fig.~\ref{fig1}(c)) corresponds to the resolved sideband limit, and is optimal to generate entanglement in the system~\cite{JieLi18,JL19NJP,JieLi20,QST23}. 

Since the dynamics of the system is fully linearized and the input noises are Gaussian, the state of the system at any given time is Gaussian, which can be characterized by an $8\times8$ covariance matrix (CM) $\mathcal{V}$, with its entries defined as $\mathcal{V}_{ij}=\langle u_i(t)u_j(t')+u_j(t')u_i(t)\rangle/2$ ($i,j=1,2,...,8$). To get the steady-state CM $\mathcal{V}$, one can directly solve the Lyapunov equation
\begin{equation}
		A \mathcal{V}+\mathcal{V} A^{T}=-D,
\end{equation}
where the diffusion matrix $D=\mathrm{diag}[\gamma_a,\gamma_a, \kappa_c, \kappa_c,0,\gamma_b(2N_b+1),\kappa_m(2N_m+1), \kappa_m(2N_m+1)]$, which is defined by $D_{ij}\delta(t-t')=\langle n_i(t)n_j(t')+n_j(t')n_i(t) \rangle/2$. In obtaining $D$, we take $N_{a(c)} \simeq 0$ due to their high mode frequencies. When the CM $\mathcal{V}$ is achieved, we can then quantify the entanglement between any two subsystems using the logarithmic negativity $E_N$~\cite{adesso}, defined as
\begin{equation}
	E_N=\textup{max}\left[ 0,-\textup{ln}(2\eta^{-})\right],
\end{equation}
where $\eta^{-} \equiv 2^{-1/2}\, \big[ \Sigma - \big(\Sigma^2 - 4\, \textup{det}\, {\cal V}_{4} \big)^{1/2}\big]^{1/2}$, and ${\cal V}_{4}=\big[{\cal V}_e, {\cal V}_{ef}; {\cal V}_{ef}^{\rm T}, {\cal V}_f \big]$ is the $4\times4$ CM of the bipartite system involving mode $e$ and $f$ ($e,f=a,c,m,b$), with ${\cal V}_e, {\cal V}_f$ and ${\cal V}_{ef}$ being the $2\times2$ blocks of ${\cal V}_{4}$, and $\Sigma \equiv \textup{det}\,{\cal V}_e+\textup{det}\, {\cal V}_f-2\textup{det}\,{\cal V}_{ef}$. 
In our highly hybrid system, tripartite entanglement may also be present. We adopt the minimum residual contangle $R_{\tau}^{\mathrm{min}}$ to quantify the tripartite entanglement, which is defined as~\cite{adesso2}
\begin{equation}
	R_{\tau}^{\mathrm{min}}\equiv \mathrm{min}[R_{\tau}^{\alpha|\beta\gamma},R_{\tau}^{\beta|\alpha\gamma},R_{\tau}^{\gamma|\alpha\beta}],
\end{equation}
where $\alpha$, $\beta$, $\gamma$ denote any three modes of the system, and $R_{\tau}^{\alpha|\beta\gamma}\equiv C_{\alpha|\beta\gamma}-C_{\alpha|\beta}-C_{\alpha|\gamma}\geq 0$ is the residual contangle, with $C_{u|v}$ ($v$ contains one or two modes) being the contangle of the $u$ and $v$ subsystems and defined as the squared logarithmic negativity~\cite{JieLi18}. A nonzero $R_{\tau}^{\mathrm{min}}>0$  indicates the presence of genuine tripartite entanglement of the corresponding three modes.

\begin{figure}[t] 
		\centering
		\includegraphics[width=1\linewidth]{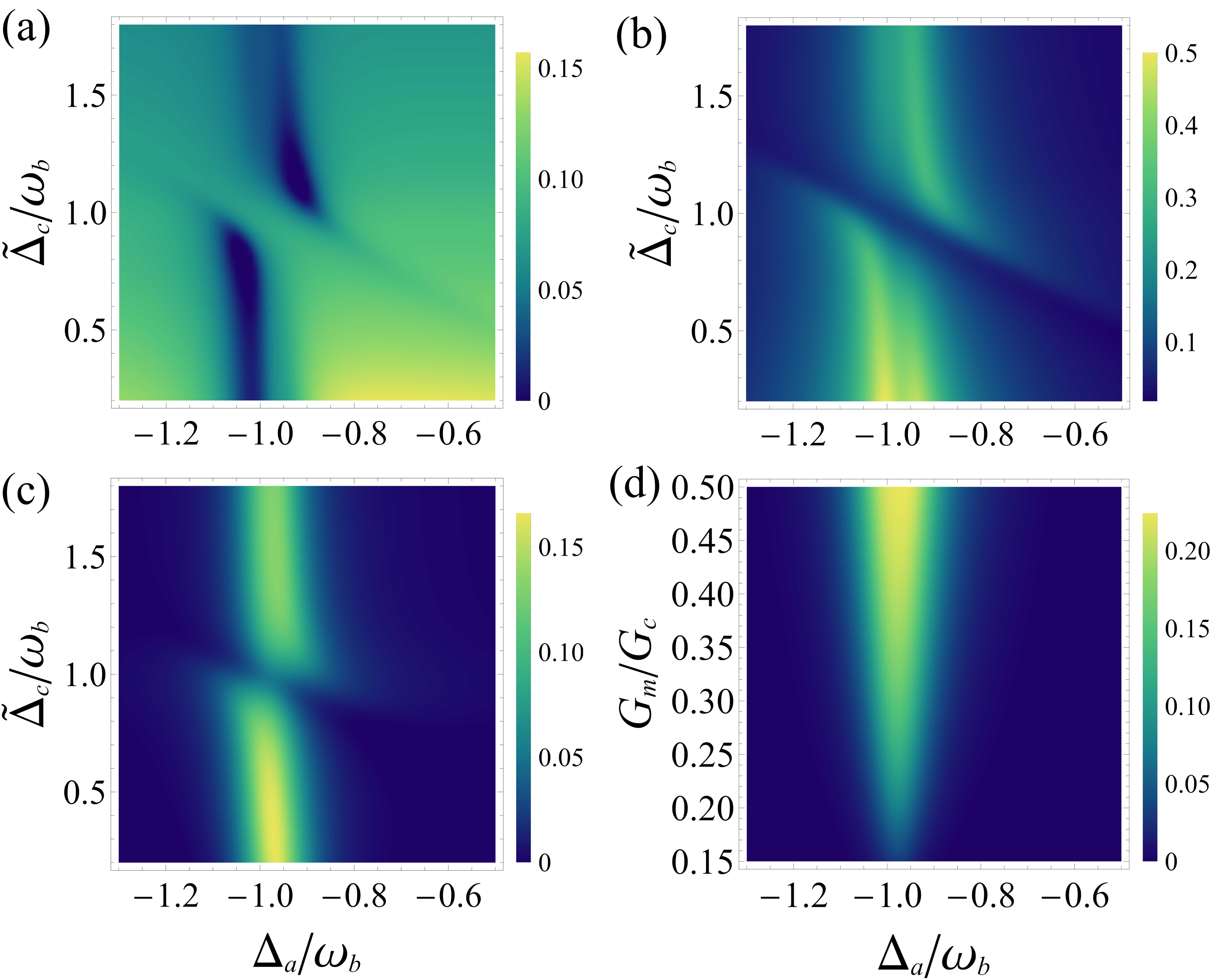}
		\caption{Density plot of steady-state (a) photon-phonon entanglement $E_{cb}$, (b) atom-phonon entanglement  $E_{ab}$, and (c) atom-magnon entanglement $E_{am}$ versus  detunings $\Delta_a$ and $\tilde{\Delta}_c$ (in units of $\omega_b$). (d) Atom-magnon entanglement $E_{am}$ versus $\Delta_a$ and magnomechanical coupling strength $G_m$ (in units of $G_c$). We take $G_c/2\pi=8$ MHz and $\tilde{\Delta}_m=\omega_b$ in all plots, $\tilde{\Delta}_c=0.5\omega_b$ in (d), and $G_m/2\pi=2.5$ MHz in (a)-(c). See text for the other parameters.} 
		\label{fig2}
\end{figure}

The mechanical mode has the lowest resonance frequency in the system, typically in megahertz~\cite{bridge,16SciAdv,JieLi18,PRX21,RCShen2022}, and is highly thermally populated even at cryogenic temperatures. Therefore, cooling the mechanical mode close to the ground state is a requisite for preparing quantum states in the system~\cite{JieLi18}.  To achieve this, we drive the cavity with a red-detuned laser, which activates the optomechanical anti-Stokes scattering and can significantly cool the mechanical mode in the resolved sideband limit $ \omega_b \gg \kappa_c$~\cite{RMP2014}. We use a relatively strong laser power, which yields a strong optomechanical coupling $G_c$ and breaks the weak-coupling condition $G_c\ll \omega_b$ for taking the rotating-wave (RW) approximation to obtain the optimal beam-splitter interaction $\propto c^\dag b + c b^\dag$ ($b\,{=}\,\frac{q+ip}{\sqrt{2}}$) for cooling. The counter-RW terms $\propto c^\dag b^\dag + c b$, corresponding to the parametric down-conversion (PDC) interaction, then start to play the role and generate the optomechanical entanglement~\cite{DV07PRL}, as shown in Fig.~\ref{fig2}(a). The anti-crossing around $\tilde{\Delta}_c=\omega_b$ in the figure is a signature of the strong coupling. The entanglement can be distributed to the atomic system when the atomic frequency matches the Stokes sideband, i.e., $\Delta_a=-\omega_b$~\cite{DV08} (c.f. Fig.~\ref{fig1}(c)), giving rise to the atom-phonon entanglement, as confirmed by Fig.~\ref{fig2}(b). A similar mechanism has been adopted to prepare entangled states in CMM systems~\cite{JieLi18,JL19NJP,JieLi20}. By further driving the magnon mode with a red-detuned microwave field with detuning $\tilde{\Delta}_m=\omega_b$, the magnomechanical anti-Stokes scattering is activated, which realizes the magnon-phonon state-swap operation. As a result, the atom-phonon entanglement is further distributed to the magnonic system, yielding stationary atom-magnon entanglement, as seen in Fig.~\ref{fig2}(c).  This entanglement transfer becomes efficient only when the magnomechanical coupling is sufficiently strong. This is clearly shown in Fig.~\ref{fig2}(d) that the atom-magnon entanglement $E_{am}$ increases as the grow of the coupling $G_m$, which increases the magnon-phonon state-swap efficiency.

In view of the whole process, from the perspective of entanglement as a finite quantum resource, it is originally generated in the optomechanical system, and then distributed to the atom-phonon system via the cavity-atom linear coupling, and further to the atom-magnon system via the phonon-magnon state-swap interaction. The complementary distributions of the entanglement in Figs.~\ref{fig2}(a), \ref{fig2}(b) and \ref{fig2}(c) are a clear sign of such an entanglement transfer process. This feature was also observed in other multipartite systems~\cite{JieLi18,Fanarxiv}.
	
In plotting Fig.~\ref{fig2}, we have used feasible parameters~\cite{bridge,16SciAdv,JieLi18,PRX21,RCShen2022,QST23}: $\omega_m/2\pi=10$ GHz, $\omega_b/2\pi=40$ MHz, $\lambda_L=852$ nm (optical wavelength),  $\kappa_m/2\pi=1$ MHz, $\gamma_a=\kappa_m$, $\kappa_c=2 \kappa_m$, $\gamma_b/2\pi=10^2$ Hz, $G_c/2\pi=8$ MHz, $g_N/2\pi=8$ MHz, $G_m/2\pi=2.5$ MHz, $\tilde{\Delta}_m=\omega_b$, and $T=10$ mK. The strong optomechanical coupling $G_c/2\pi=8$ MHz can be achieved with a laser power $P_L\simeq 5.5$ mW for $g_c/2\pi=1$ kHz at detunings $\tilde{\Delta}_c\simeq 0.5\omega_b$ and $\Delta_a=-\omega_b$. The magnomechanical coupling $G_m/2\pi=2.5$ MHz corresponds to a microwave drive power $P_0\simeq 1.44$ mW for a $5\times2\times1$ $\mu m^3$ YIG bridge (approximated as a cuboid) with $g_m/2\pi=20$ Hz~\cite{QST23}.

\begin{figure}[t]
		\includegraphics[width=\linewidth]{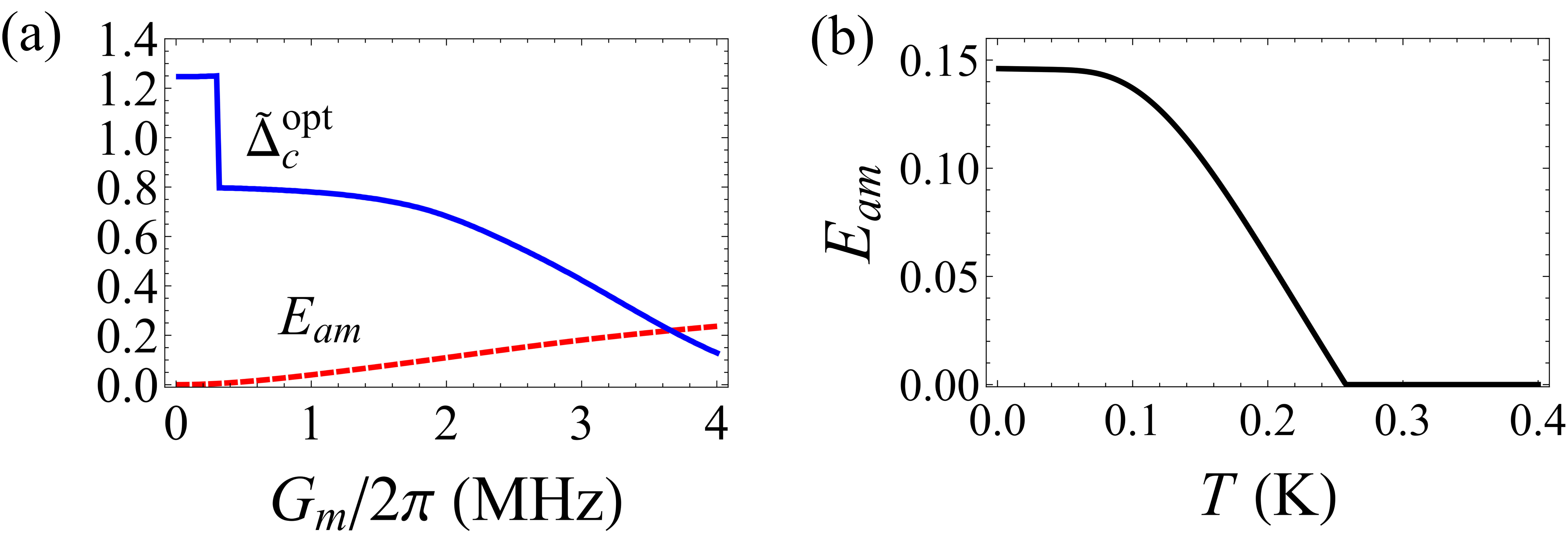}
		\caption{(a) Stationary atom-magnon entanglement $E_{am}$ (dashed) and the corresponding optimal cavity detuning $\tilde{\Delta}_c^{\rm{opt}}$ (solid, in unit of $\omega_b$ ) versus $G_m$. (b) $E_{am}$ versus bath temperature $T$. We take $\tilde{\Delta}_m=-\Delta_a=\omega_b$ in both plots, and $\tilde{\Delta}_c=0.5\omega_b$ and $G_m/2\pi=2.5$ MHz in (b). The other parameters are the same as in Fig.~\ref{fig2}(c).}
		\label{fig3}
\end{figure}

In our system, since the cavity and magnon modes are simultaneously driven by red-detuned laser and microwave fields (Fig.~\ref{fig1}(c)), both the opto- and magnomechanical anti-Stokes scatterings are present in the system and these two mechanical cooling channels may compete when the drive fields are sufficiently strong. Such competition effect is revealed in Fig.~\ref{fig3}(a), where we plot the atom-magnon entanglement $E_{am}$ and the corresponding optimal cavity-laser detuning $\tilde{\Delta}_c^{\rm{opt}}$ versus the magnomechanical coupling $G_m$. When $G_m$ is small, the optomechanical cooling channel is dominant and thus the optimal detuning $\tilde{\Delta}_c^{\rm{opt}}$ (for the entanglement) is around $\omega_b$, at which the cooling efficiency is maximized~\cite{RMP2014}. As $G_m$ grows, the magnon-phonon state-swap efficiency increases yielding an increasing $E_{am}$ (dashed line). Besides, the contribution of the magnomechanical anti-Stokes scattering in mechanical cooling increases, which leads to a decreasing optimal detuning $\tilde{\Delta}_c^{\rm{opt}}$ (solid line). Such a shift of the optimal detuning implies that the strength of the optomechanical beam-splitter interaction reduces (its role in mechanical cooling is gradually replaced by the magnomechanical beam-splitter interaction), while the strength of the optomechanical PDC interaction increases.  The optimal detuning $\tilde{\Delta}_c^{\rm{opt}}$ is the result of the trade-off among the magnomechanical cooling, the optomechanical cooling and PDC. It is worth noting that the steep drop of $\tilde{\Delta}_c^{\rm{opt}}$ as $G_m$ grows in Fig.~\ref{fig3}(a) is due to the anti-crossing associated with the strong coupling (c.f. Fig.~\ref{fig2}),  corresponding to the sudden transition of $\tilde{\Delta}_c^{\rm{opt}}$ between two branches near the anti-crossing.  The generated stationary macroscopic atom-magnon entanglement can exist for bath temperature up to hundreds of millikelvin, as show in Fig.~\ref{fig3}(b), under a moderate coupling $G_m/2\pi=2.5$ MHz.

\begin{figure}[b] 
	\centering
	\includegraphics[width=1\linewidth]{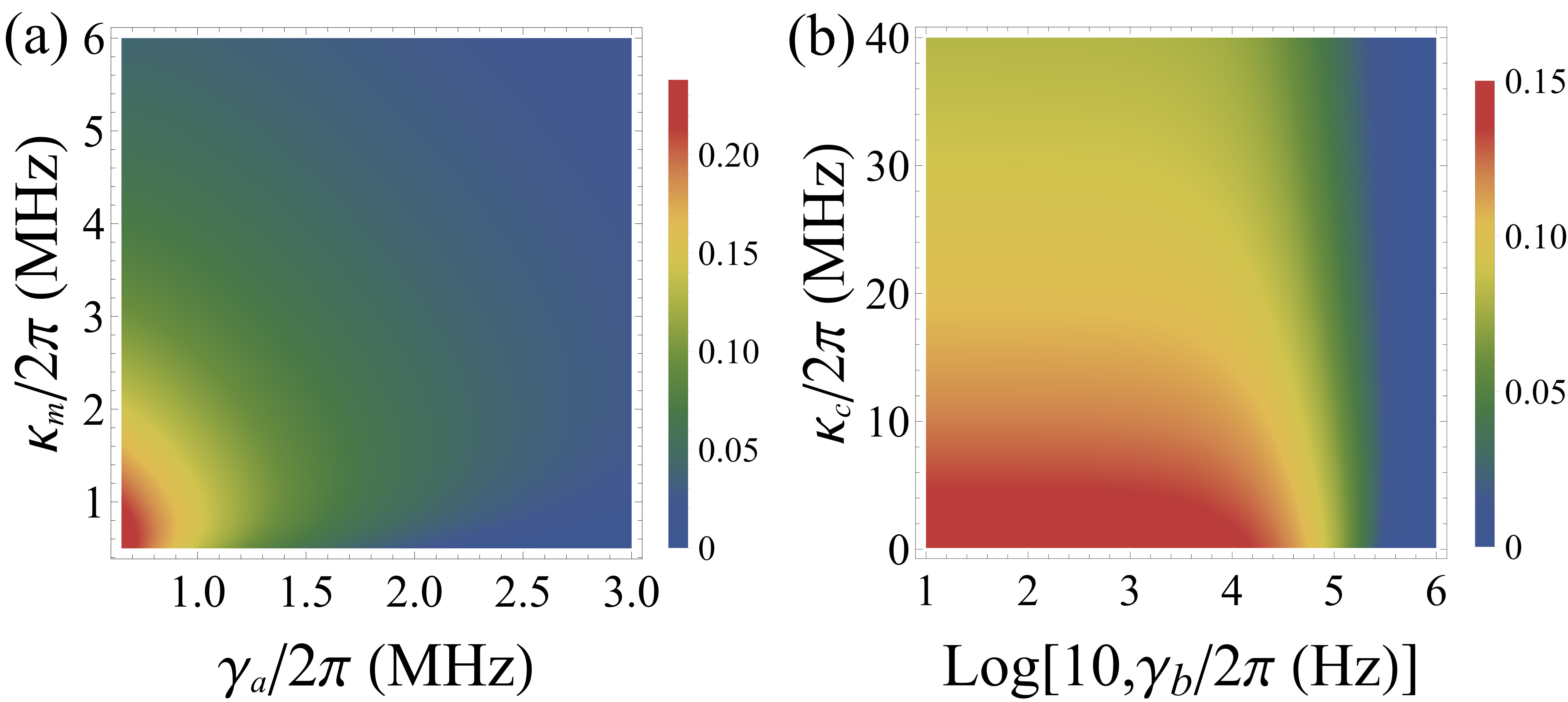}
	\caption{Stationary atom-magnon entanglement $E_{am}$ versus dissipation rates (a) $\gamma_{a}$ and $\kappa_{m}$; and (b) $\gamma_b$ and $\kappa_c$. We take bath temperature $T=10$ mK and the other parameters are the same as in Fig.~\ref{fig3}(b).} 
	\label{fig4}
\end{figure}

\begin{figure}[t] 
	\centering
	\includegraphics[width=1\linewidth]{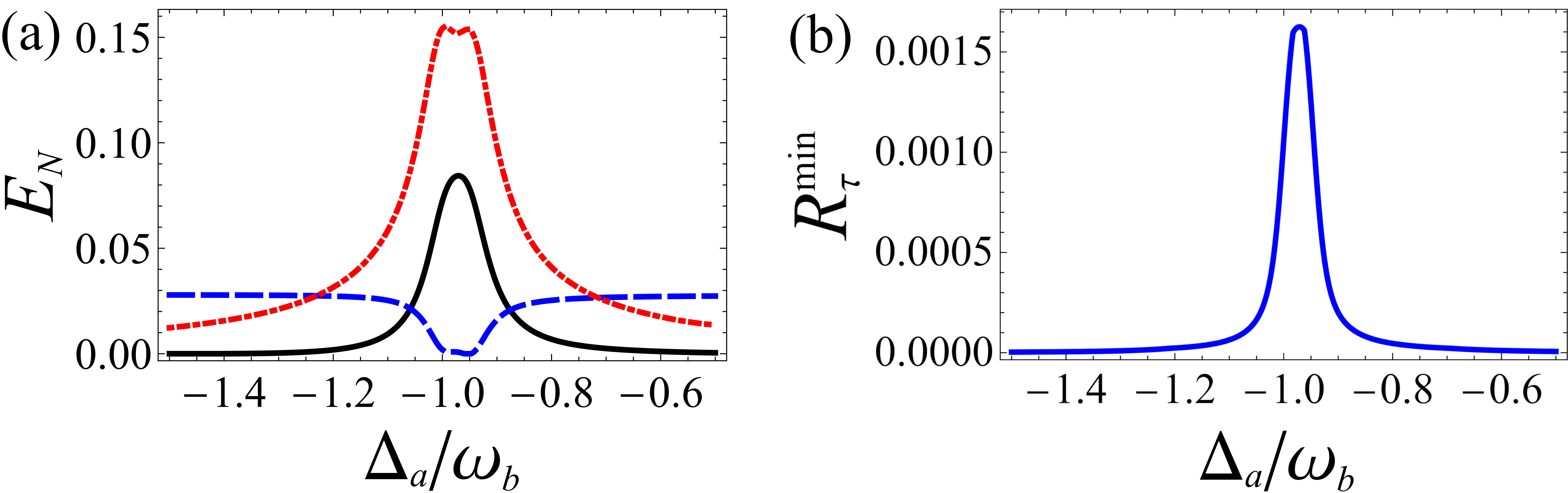}
	\caption{(a) Bipartite entanglements $E_{ab}$ (dot-dashed), $E_{bm}$ (dashed), and $E_{am}$ (solid) versus $\Delta_a$. (b) Tripartite atom-magnon-phonon entanglement in terms of the minimum residual contangle $R^{\rm{min}}_{\tau}$ versus $\Delta_a$. We take $G_c/2\pi=2.3$ MHz and $T=10$ mK. The other parameters are the same as in Fig.~\ref{fig3}(b).} 
	\label{fig5}
\end{figure}

We now study the effect of the dissipations of each mode on the entanglement. In Fig. \ref{fig4}, we plot the stationary atom-magnon entanglement versus four dissipation rates of the system. Clearly, the entanglement is robust against all the dissipation rates and exists in a wide range of the dissipation rates. It is, however, more sensitive to the atomic and magnonic decay rates $\gamma_{a}$ and  $\kappa_m$, and the entanglement is present only when $\gamma_{a}, \kappa_m < 2\pi \times 10$ MHz.  By contrast, it is much more robust towards the cavity decay rate $\kappa_c$ and the mechanical damping rate $\gamma_{b}$. For $\kappa_c$ being up to tens of MHz within the resolved-sideband limit $\kappa_c<\omega_b$, and $\gamma_b$ up to $\sim 2\pi \times 4\times10^5$ Hz (corresponding to a quality factor $Q_b\sim100$), the entanglement is still present.
Note that Fig. \ref{fig4} is plotted using the exact drift matrix $A$ in Eq. \eqref{Eqmatrix}, without assuming real couplings $G_m$ and $G_c$. The exact form of the drift matrix $A$ with generally complex couplings $G_m$ and $G_c$ is provided in Ref. \cite{note}.

Apart from the presence of many bipartite entanglements in the system, as displayed in Fig.~\ref{fig2}, the complex quadripartite system may exhibit multipartite entanglement.  Despite the atom-cavity-phonon entanglement that has been studied in a simpler configuration \cite{DV08}, here we reveal that considerable tripartite entanglement is shared among the three massive subsystems, i.e., atoms, magnons and phonons, as illustrated in Fig. \ref{fig5}.  The simultaneous existence of all bipartite entanglements $E_{ab}$, $E_{bm}$ and $E_{am}$ and genuine tripartite entanglement $R^{\rm{min}}_{\tau}>0$ in the atom-magnon-phonon system is clear evidence for the strong quantum correlation shared by the three macroscopic systems.  There are also other forms of tripartite entanglement in the system, but their degree of entanglement is small. Therefore, those results will not be presented here.  

Lastly, we discuss the validity of our model and provide a strategy to detect and verify the entanglement.  The bosonic description for the spin systems is valid in the low-excitation limit, i.e., $\langle m^\dagger m \rangle \ll 2Ns$ and $\langle a^\dagger a \rangle \ll N_a$. Under the parameters of Fig.~\ref{fig3}(b), we obtain $\langle m^\dagger m \rangle = 7.7\times 10^9 \ll 2Ns=2.1 \times 10^{11}$ and $\langle a^\dagger a \rangle = 1.3 \times 10^6 \ll N_a=6.4\times 10^7$. In estimating $N_a$, we take $g_a/2\pi=10^3$ Hz~\cite{DV08}. Clearly, the low-excitation condition is well fulfilled. For the entanglement detection, the magnon state can be accessed by coupling to a microwave cavity that is driven by a {\it weak} probe field. Due to the magnon-cavity beam-splitter coupling~\cite{naka,zou}, the magnon state can be read out in the cavity output field. By homodyning the microwave output field, two quadratures of the magnon mode can be measured. Similarly, the atomic polarization quadratures can be measured by coupling to an additional optical cavity that is driven by a weak laser field. Having measured the magnonic and atomic quadratures, one can then build the CM, based on which the atom-magnon entanglement is computed~\cite{Palomaki,mechanical3,DV07PRL}.

In conclusion, we present a protocol to prepare stationary entangled states of two macroscopic systems, an atomic ensemble and ferrimagnetic magnons, in an opto-magnomechanical system. The atom-magnon entanglement is established as a result of the combination of opto- and magnomechanical cooling and optomechanical PDC interactions, and the entanglement distribution among different subsystems. A competition mechanism between two opto- and magnomechanical cooling channels is revealed.  We further confirm the presence of genuine tripartite entanglement in the atom-magnon-phonon system, where all three subsystems are massive.  
The protocol may find potential applications in preparing macroscopic quantum states, given that macroscopic entangled states of two atomic ensembles~\cite{EP01} and of an atomic ensemble and a mechanical oscillator ~\cite{EP21} have been successfully generated.  The atom-magnon entanglement can also be used to prepare microwave-optics entanglement by coupling atoms (magnons) to an optical (a microwave) field, which is of particular importance in building a hybrid quantum network~\cite{Qiu}. 

This work has been supported by National Key Research and Development Program of China (Grant No. 2022YFA1405200) and National Natural Science Foundation of China (No. 92265202).


\begin{thebibliography}{99}
\bibitem{note}
The exact form of the drift matrix $A$ is given by
\begin{equation*}
		A=\begin{pmatrix}
		-\gamma_a & \Delta_a & 0 & g_N & 0 & 0 & 0 & 0\\
		-\Delta_a & -\gamma_a & -g_N & 0 & 0 & 0 & 0 & 0\\
		0 & g_N & -\kappa_c & \tilde{\Delta}_c & -\mathrm{Im}G_c & 0 & 0 & 0\\
		-g_N & 0 & -\tilde{\Delta}_c & -\kappa_c & \mathrm{Re}G_c & 0 & 0 & 0\\
		0 & 0 & 0 & 0 & 0 & \omega_b & 0 & 0\\
		0 & 0 & \mathrm{Re}G_c & \mathrm{Im}G_c & -\omega_b & -\gamma_b & -\mathrm{Re}G_m & -\mathrm{Im}G_m\\
		0 & 0 & 0 & 0 & \mathrm{Im}G_m & 0 & -\kappa_m & \tilde{\Delta}_m\\
		0 & 0 & 0 & 0 & -\mathrm{Re}G_m & 0 & -\tilde{\Delta}_m & -\kappa_m
	\end{pmatrix} ,
\end{equation*}
where the effective opto- and magnomechanical coupling strengths are redefined as $G_c=\sqrt{2}g_c\langle c\rangle$ and $G_m=\sqrt{2}g_m\langle m\rangle$, because of the complex  averages $\langle c\rangle$ and $\langle m\rangle$ as provided in Eq.\eqref{mean}. Figure \ref{fig4} is plotted with fixed amplitudes of the couplings $|G_c|/2\pi=8$ MHz and $|G_m|/2\pi=2.5$ MHz.
\end{thebibliography}
\end{document}